\documentclass{article}
\usepackage{spconf,amsmath,graphicx,hyperref,xcolor,booktabs,tabularx,multirow}
\usepackage{enumitem}
\hypersetup{hidelinks}
\AddToHook{cmd/thebibliography/after}{\setlength{\itemsep}{0pt}}


\title{From Speech to Editable Concepts: Probing Emotion Recognition with Concept Bottleneck Models}
\name{Hezhao Zhang \qquad Thomas Hain}
\address{School of Computer Science, University of Sheffield\\
Sheffield, United Kingdom}
\begin{document}
\raggedbottom
\maketitle
\begin{abstract}
Speech emotion recognition (SER) is the task of assigning emotion labels to utterances. Early systems relied on acoustic features, whereas recent approaches combine multiple modalities, most commonly speech and text. Still, performance remains poor on many datasets. Large language models (LLMs) have therefore attracted interest for SER, as they can process diverse inputs jointly with instructions.
However, direct audio input raises questions of explainability. To address similar questions in image classification, concept bottleneck models were introduced. This work adapts concept bottlenecks to SER to examine how individual predictions depend on transcripts, acoustic descriptions and speaker attributes.
Experiments test three LLMs on CREMA-D, IEMOCAP and MELD, with concepts extracted by separate tools. On scripted corpora, LLMs are strongly biased towards the transcript in the zero-shot setting, which lowers Macro-F1 from 27.8 to 5.8 on CREMA-D. Fine-tuning removes this bias, and the transcript raises Macro-F1 from 41.8 to 45.1. Removing speech rate changes 48\% of Neutral predictions to Disgust on CREMA-D; removing intensity level on MELD changes predictions despite little change in Macro-F1. These findings show that aggregate performance changes alone do not capture the effects of concept removal on individual predictions.

\end{abstract}
\section{Introduction}
\label{sec:intro}

Speech emotion recognition (SER) predicts the emotion label of an utterance from the speech signal \cite{Chul_Min_Lee_2005}.
Early systems trained classifiers on a wide range of carefully selected acoustic features such as pitch, energy and spectral descriptors \cite{schullerINTERSPEECH2009Emotion2009,eybenGenevaMinimalisticAcoustic2016}.
Later, neural networks learned acoustic representations directly from the raw waveform \cite{trigeorgisAdieuFeaturesEndtoend2016}, and self-supervised speech representations turned out to be better as classifier input \cite{pepinoEmotionRecognitionSpeech2021}.
However, besides acoustic information, speech also carries lexical content \cite{Chul_Min_Lee_2005,Yoon_2018}, and subsequent systems therefore combined the speech representations with text features from the transcript \cite{Sun_2023}.
More recently, SER research has adopted large audio-language models (LALMs), models that extend large language models (LLMs) with an audio encoder and predict the emotion label directly from the audio given a text instruction \cite{wang-etal-2024-blsp,maiAASLLMAcousticallyAugmented2025}.
Despite these improvements recognition performance remains low on many corpora \cite{maEmoBoxMultilingualMulticorpus2024,zhangVoxEmoBenchmarkingSpeech2026}.
Moreover, with the direct audio input, the LALM gives no indication of which information it relies on, so its decisions cannot be examined.

Several works make use of acoustic information in textual form. 
EmotionThinker \cite{wangEmotionThinkerProsodyAwareReinforcement2026} makes its reasoning visible by keeping the audio as input and generating a reasoning text that describes prosody before giving the label.
SpeechCueLLM \cite{wuSilentLettersAmplifying2025} and VowelPrompt \cite{wangVowelPromptHearingSpeech2026} make the input tangible by extracting acoustic descriptions from the audio and providing them, together with the transcript, to a text LLM.
However, retraining without individual descriptions or swapping descriptions between utterances \cite{wangVowelPromptHearingSpeech2026} does not identify which description a fixed predictor's decision depends on; generated reasoning may not faithfully reflect how the model reached its prediction \cite{NEURIPS2023_ed3fea90}.

Identifying which information affects a prediction requires the ability to control the model input and knowing what such input represents.
A model that uses human-readable form as input is therefore desirable. 
Image classification faced the same problem, and concept bottleneck models (CBMs) were introduced in response \cite{kohConceptBottleneckModels2020}.
Here an image is first mapped to human-readable concepts, and the label is predicted from these alone. Hence every prediction can be traced to specific concepts and can be changed by editing them.
Shin et al. \cite{shinCloserLookIntervention2023} used such edits to examine how predictions respond to individual concepts.

We apply the CBM idea to SER: separate extractors describe each utterance by concepts that are human readable (Section~\ref{sec:method}). Then an LLM predicts the emotion label from these concepts, enabling single-concept removal with the predictor fixed to examine which class predictions change.  Here concepts are the transcript, pitch, intensity and speech rate, as well as speaker age and gender. Together these describe the words spoken, intonation and speaker characteristics.

This work studies how the concepts affect emotion decisions: (1) what does each concept group contribute to recognition across corpora, before and after fine-tuning? (2) which acoustic concepts do the decisions depend on, and how does this dependence vary across corpora?
The first question is addressed by comparing predictors that receive different concept groups (Section~\ref{sec:composition-results}).
The second by removing single acoustic concepts from the input of a predictor fine-tuned on all concepts (Section~\ref{sec:replacement-results}), with both analyses run on CREMA-D \cite{caoCREMADCrowdSourcedEmotional2014}, IEMOCAP \cite{bussoIEMOCAPInteractiveEmotional2008} and MELD \cite{poriaMELDMultimodalMultiParty2019} with three LLMs (Section~\ref{sec:setup}).

\section{Speech Concept Bottleneck}
\label{sec:method}

A concept bottleneck model operates in two cascaded stages \cite{kohConceptBottleneckModels2020}:
A concept extractor $g$ maps an input $x$ to a set of concepts $c$, and a predictor $f$ predicts the label from $c$ only,
\begin{equation}
\hat{y} = f(c), \qquad c = g(x).
\label{eq:cbm}
\end{equation}
Concepts are human-specified, interpretable properties of the input, whose predicted values form the input to $f$. A concept intervention changes one concept while $f$ is held fixed and observes the change in $\hat{y}$ \cite{kohConceptBottleneckModels2020,shinCloserLookIntervention2023}.
This work adopts this framework for SER: the input $x$ is an utterance, the concepts $c$ are properties of speech expressed as text, and the predictor $f$ is an LLM that processes this text to produce one emotion label from a fixed set.

The concepts used are organised into three groups, each related to how emotion is expressed in speech (see Table~\ref{tab:prompt}).
The transcript represents content and gives the words spoken, which carry emotion information through their meaning. The acoustic concepts are pitch, intensity and speech rate, three established acoustic correlates of emotion \cite{schererVocalCommunicationEmotion2003,eybenGenevaMinimalisticAcoustic2016,wuSilentLettersAmplifying2025}.
In addition, the expression and perception of emotion in speech vary with the speaker's age and gender \cite{senAgeDifferencesVocal2018,lausenGenderDifferencesRecognition2018}.
To account for these differences, the speaker concepts include the age and gender of each speaker. Each group is obtained from a separate off-the-shelf extractor that was not trained on emotion labels (see Sec.~\ref{sec:extraction}).

\begin{table}[t]
\caption{Prompt template with all three concept groups. Unselected groups are omitted. Volume denotes intensity and Predicted sex the speaker's gender.}
\label{tab:prompt}
\centering
\fontsize{9}{11}\selectfont
\providecommand{\ph}[1]{$\langle$\textit{#1}$\rangle$}
\begin{tabularx}{\columnwidth}{@{}>{\raggedright\arraybackslash}X@{}}
\toprule
Predict the emotion expressed in the utterance from the provided information. \\
\addlinespace
Transcript: \ph{transcript} \\
\addlinespace
Pitch level: \ph{5 levels from very low to very high}. \\
Pitch variation: \ph{5 levels from very low to very high}. \\
Volume level: \ph{5 levels from very low to very high}. \\
Volume variation: \ph{5 levels from very low to very high}. \\
Speech rate: \ph{5 levels from very slow to very fast}. \\
\addlinespace
Estimated age: \ph{6 bands from under 20 to 60+}. \\
Predicted sex: \ph{Female or Male}. \\
\addlinespace
Allowed labels: \ph{label set} \\
\addlinespace
Return exactly one label from the allowed labels. Do not provide an explanation. \\
\bottomrule
\end{tabularx}
\end{table}

\begin{table*}[!t]
\begingroup
\fontsize{10}{12}\selectfont
\caption{Test Macro-F1 (\%) of each combination of concept groups, zero-shot and fine-tuned.}
\label{tab:rq1-g}
\centering
\setlength{\tabcolsep}{3pt}
\renewcommand{\arraystretch}{1.0}
\setbox0=\vbox\bgroup
\begin{tabular*}{\textwidth}{@{\extracolsep{\fill}}llrrrr@{\hspace{40pt}}rrrr@{}}
\toprule
 &  & \multicolumn{4}{c}{\textbf{Zero-shot}} & \multicolumn{4}{c}{\textbf{Fine-tuned}} \\
\cmidrule(lr){3-6}\cmidrule(lr){7-10}
Dataset & Model & T & A & TA & TAP & T & A & TA & TAP \\
\midrule
\textbf{CREMA-D} & Qwen2.5 & 4.26 & \textbf{27.88} & 5.87 & 7.13 & 11.74 & 43.03 & 44.21 & \textbf{45.50} \\
 & Qwen2.5-Omni & 4.26 & \textbf{24.24} & 4.79 & 4.79 & 11.28 & 42.18 & 44.51 & \textbf{45.57} \\
 & Llama 3.1 & 4.26 & \textbf{25.01} & 16.80 & 13.80 & 11.15 & 41.87 & 45.10 & \textbf{45.37} \\
\midrule
\textbf{IEMOCAP} & Qwen2.5 & 46.11 & 36.49 & 51.71 & \textbf{52.15} & 69.90 & 45.49 & \textbf{74.37} & 74.16 \\
 & Qwen2.5-Omni & 48.43 & 31.41 & \textbf{51.99} & 51.83 & 70.26 & 45.17 & \textbf{74.97} & 74.58 \\
 & Llama 3.1 & 52.32 & 32.35 & \textbf{54.49} & 53.14 & 70.49 & 44.38 & \textbf{74.24} & 74.05 \\
\midrule
\textbf{MELD} & Qwen2.5 & \textbf{36.16} & 13.77 & 33.83 & 33.13 & 37.20 & 13.51 & 37.03 & \textbf{39.02} \\
 & Qwen2.5-Omni & \textbf{33.40} & 11.80 & 32.70 & 32.76 & 37.19 & 13.71 & \textbf{38.48} & 37.98 \\
 & Llama 3.1 & \textbf{32.83} & 11.80 & 30.37 & 30.68 & 37.27 & 13.97 & \textbf{38.50} & 37.36 \\
\bottomrule
\end{tabular*}
\egroup
\typeout{RQ1-G-BODY: width=\the\wd0; height=\the\ht0; depth=\the\dp0}
\nointerlineskip\box0
\par\vspace{3pt}
\begin{minipage}{\textwidth}
\fontsize{9}{10.5}\selectfont
\raggedright T: transcript; A: acoustic concepts; P: speaker concepts, all given as text. Bold: highest mean per model and setting; ties share the mark. Fine-tuned: mean of three runs (seed SD 0.1--3.6 points); IEMOCAP: five-fold means. Direct-audio reference: Table~\ref{tab:rq1-audio}.
\end{minipage}
\endgroup

\end{table*}

\section{Implementation and Experiments}
\label{sec:setup}

\subsection{Concept Extraction and Prompting}
\label{sec:extraction}

This study aims to provide insight into how the predictor makes use of the concepts, not how they are extracted. In order to extract concepts of high quality, each is obtained independently  state of the art tools.
The transcript is produced by Qwen3-ASR~\cite{shiQwen3ASRTechnicalReport2026}. Pitch and intensity are measured with Praat~\cite{jadoulIntroducingParselmouthPython2018}, with the mean over the utterance giving the level and the standard deviation giving the variation. Speech rate is defined as the number of words in the transcript divided by the utterance duration~\cite{yildirimAcousticStudyEmotions2004}.
Each acoustic measure is discretised into five levels with quantile thresholds fitted on the training set of each dataset and fold. The prompt is constructed with a simple  description of the level (Table~\ref{tab:prompt}). The speaker concepts are derived from Vox-Profile~\cite{fengVoxProfileSpeechFoundation2025}, with the predicted age grouped into six ten-year bands from under 20 to over 60.
The predicted gender agrees with the metadata for 96.3\% of CREMA-D and 95.2\% of IEMOCAP utterances. Age metadata exist only for CREMA-D, where the predicted band is correct for only 35.9\% of utterances and within one band for 78.2\%. The prompt contains the selected concept groups and allowed labels, without specifying how concept values relate to emotions.

\subsection{Experimental Design}
\label{sec:design}

To measure what each concept group contributes, the four combinations in Table~\ref{tab:rq1-g} are evaluated with the zero-shot LLM and with a predictor fine-tuned on each combination alone.
The gain from adding a group to utterance descriptions is considered its contribution, in zero-shot setting and after fine-tuning.
To identify the relation between acoustic concepts and decisions , the predictor fine-tuned on all concepts is held fixed. The same utterance is assessed with and without a specific acoustic concept , (as defined in Sec.~\ref{sec:method} ), thus providing some indication of its relevance. 

\subsection{Experimental Setup}
\label{sec:conditions}

\textbf{Datasets.}
Three corpora are used in which emotion is carried by the words and by the voice to different degrees.
In CREMA-D \cite{caoCREMADCrowdSourcedEmotional2014}, twelve fixed neutral sentences are acted in six emotions, so only the voice carries emotion.
IEMOCAP \cite{bussoIEMOCAPInteractiveEmotional2008} holds scripted and improvised dyadic sessions performed by actors, and MELD \cite{poriaMELDMultimodalMultiParty2019} multi-party television dialogue, and in both the words carry emotion as well.
CREMA-D has six classes and speaker-disjoint splits, IEMOCAP four classes and session-wise five-fold cross-validation, and MELD seven classes and official splits.

\textbf{Models.}
Three open instruction-tuned LLMs serve as the predictor: Qwen2.5-7B-Instruct \cite{qwenQwen25TechnicalReport2025}, Qwen2.5-Omni-7B \cite{xuQwen25OmniTechnicalReport2025}, which is built on Qwen2.5-7B, and Llama-3.1-8B-Instruct \cite{grattafioriLlama3Herd2024}.
Qwen2.5-Omni is also given the audio directly as a reference condition, zero-shot and fine-tuned.

\textbf{Training.}
Fine-tuning uses LoRA \cite{huLoRALowRankAdaptation2021} with rank 16, $\alpha=32$ and dropout 0.05 on the attention and feed-forward projections.
Optimisation uses AdamW \cite{loshchilovDecoupledWeightDecay2019} with a learning rate of $2\times10^{-4}$, 10\% warm-up and linear decay, for six epochs with a batch size of 16 in BF16.
Each model is fine-tuned on each combination with three seeds and on each IEMOCAP fold.

\textbf{Evaluation.}
Performance is measured as Macro-F1 under greedy decoding, and an output that cannot be parsed as a label counts as an error.
The effect of a concept removal is the change in Macro-F1 relative to the full-input score of the same run.

\section{Results and Discussion}
\label{sec:results}
\subsection{Predictive Performance}
\label{sec:composition-results}
Table~\ref{tab:rq1-g} reports Macro-F1 for each combination of concept groups, zero-shot and after fine-tuning.
On CREMA-D, zero-shot models are strongly biased towards the transcript, which is detrimental.
As the same sentences are spoken in every emotion, the models given the transcript alone classify every utterance as Neutral, giving 4.26 Macro-F1 for all of them.
Although the acoustic concepts alone reach 27.88 for Qwen2.5, adding the transcript brings the score back to 5.87, with most predictions again Neutral, and the other two models drop in the same way.

\begin{table}[t]
\begingroup
\fontsize{10}{12}\selectfont
\caption{Qwen2.5-Omni test Macro-F1 (\%): highest-scoring concept combination from Table~\ref{tab:rq1-g} versus direct audio input.}
\label{tab:rq1-audio}
\centering
\setlength{\tabcolsep}{3pt}
\renewcommand{\arraystretch}{1.03}
\begin{tabular*}{\linewidth}{@{\extracolsep{\fill}}lrrrr@{}}
\toprule
 & \multicolumn{2}{c}{Zero-shot} & \multicolumn{2}{c}{Fine-tuned} \\
\cmidrule(lr){2-3}\cmidrule(lr){4-5}
Dataset & Concepts & Audio & Concepts & Audio \\
\midrule
CREMA-D & 24.24 & 54.95 & 45.57 & 78.67 \\
IEMOCAP & 51.99 & 69.32 & 74.97 & 82.66 \\
MELD & 33.40 & 34.91 & 38.48 & 41.07 \\
\bottomrule
\end{tabular*}
\par\vspace{3pt}
\begin{minipage}{\linewidth}
\fontsize{9}{10.5}\selectfont
\raggedright Audio fine-tuning also adapts the encoder and projector. Fine-tuned scores average three runs.
\end{minipage}
\endgroup

\vspace{-8pt} %
\end{table}

However, the fine-tuned models show the opposite pattern.
After fine-tuning, the models decide from the acoustic concepts, and Llama 3.1 reaches 41.87 with these alone.
The transcript alone still gives only 11.15.
Yet adding it to the acoustic concepts now raises the score to 45.10 rather than lowering it.
The gain comes from combining the two inputs, and the reversal holds for all three models.

On IEMOCAP and MELD, the same bias towards the transcript does no harm.
These are conversational corpora, so the words change with the emotion, and the transcript alone already recognises it, giving 46.11 for Qwen2.5 on IEMOCAP.
With the acoustic concepts alone, zero-shot Qwen2.5 scores 36.49 on IEMOCAP, and adding the transcript raises this to 51.71, where the same addition lowered the score on CREMA-D.
The acoustic concepts still improve recognition on IEMOCAP, where fine-tuning with both inputs reaches 74.37 against 69.90 with the transcript, but gains on MELD are small and inconsistent across models: the same comparison gives 37.03 against 37.20 for Qwen2.5.

Adding the speaker concepts changes the fine-tuned score by less than two points on every corpus, without a consistent direction.
These small gains may reflect both limited within-speaker variation and errors in the extracted speaker concepts.
Even at its best, the concept input stays below direct audio, and the gap depends on the corpus.
Fine-tuned Qwen2.5-Omni reaches 45.57 on CREMA-D from concepts but 78.67 from audio, while the gap shrinks to 7.69 on IEMOCAP and 2.59 on MELD (Table~\ref{tab:rq1-audio}).
The gap is largest where the emotion lies in the voice and smallest where it lies in the words.

\begin{figure}[t]
\centering
\includegraphics[width=\columnwidth]{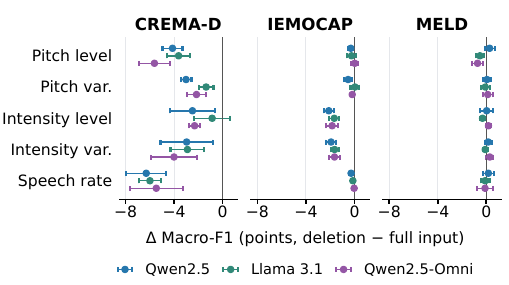}
\caption{Changes in Macro-F1 relative to the fine-tuned TAP scores in Table~\ref{tab:rq1-g}, after removing one acoustic concept while keeping the predictor fixed. Points show three-run means and error bars show $\pm1$ standard deviation. IEMOCAP folds are averaged equally within each run.}
\label{fig:single-attribute-deletion}
\end{figure}

\subsection{Concept Interventions}
\label{sec:replacement-results}
Fig.~\ref{fig:single-attribute-deletion} shows the change in Macro-F1 when one acoustic concept is removed from the input of the predictor fine-tuned on all concepts.
How much a model depends on single acoustic concepts, and on which, follows the corpus.
On CREMA-D, where the emotion lies in the acoustic concepts, removing any of the five lowers Qwen2.5's Macro-F1, by 2.49 to 6.30.
The largest loss comes from speech rate.
In this corpus, speech rate separates Neutral from Disgust: Neutral utterances tend to be fast, with 37.0\% in the fastest level and 6.5\% in the slowest, whereas Disgust utterances tend to be slow, with 11.9\% in the fastest level and 29.4\% in the slowest.
IEMOCAP, where the transcript carries more of the emotion, loses less than 0.6 for three of the five concepts.
The exceptions are intensity level and variation, at 2.12 and 1.93, and intensity is again the concept that separates a class pair: 35.0\% of Sad utterances have the lowest intensity level against 13.8\% of Neutral.
On MELD, no removal changes the score by more than 0.3.
There, the emotions are spoken at nearly the same average speech rate, intensity and pitch, so the acoustic concepts separate them little: even the two classes furthest apart on speech rate, Happy and Disgust, differ by 0.7 of a level.
The pattern holds for all three models, although for Qwen2.5-Omni pitch level and speech rate cost about the same on CREMA-D.

\begin{figure}[t]
\centering
\includegraphics[width=\columnwidth]{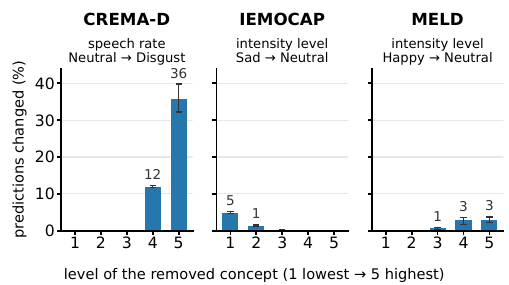}
\caption{Predictions that change when one concept is removed, as a share of all predictions of the first class (\%), by the utterance's level of the removed concept. Bars pool the three models and whiskers span the per-model values.}
\label{fig:transition}
\end{figure}

Fig.~\ref{fig:transition} removes the concept with the largest loss on CREMA-D and IEMOCAP, speech rate and intensity level, and intensity level on MELD for comparison, and follows the predictions that change.
For each corpus it shows the label that changes most often, as a share of its predictions, and the label it turns into, split by the utterance's level of the removed concept.
These changes come almost entirely from the two highest or the two lowest levels of the removed concept, and few from the middle levels.
Without speech rate, 48\% of the Neutral predictions on CREMA-D become Disgust, all from the two fastest levels.
Removing speech rate changes these Neutral predictions to Disgust, with most transitions occurring at high speech-rate levels.
Without intensity level, 6\% of the Sad predictions on IEMOCAP become Neutral, all from the two lowest levels.
Angry and Happy predictions move to Neutral in the same way, from the loudest levels.
On MELD, 7\% of the Happy predictions become Neutral, mostly from the two loudest levels.
Predictions also move from Neutral to Happy, despite little change in Macro-F1. The three models change the same levels and differ only in the size of the CREMA-D top level, from 32 to 40.
These results reveal corpus-dependent prediction sensitivity to concept removal.

\vspace*{-2mm}

\section{Conclusion}
\label{sec:conclusion}
\vspace*{-2mm}
The concept bottleneck framework was adopted for SER: each utterance is represented by transcript, acoustic and speaker concepts, and an LLM predicts the emotion from these concepts alone. Experiments were conducted in zero-shot and fine-tuned settings. In the zero-shot setting, the transcript concept dominates outcomes, which is found to be detrimental on scripted corpora, where the transcript is neutral.
After fine-tuning, the contributions of transcripts and acoustic concepts vary across corpora. Concept-based prediction still underperforms direct audio input, with the largest gap on CREMA-D, where the fixed transcripts carry no emotion information. Removing individual acoustic concepts from a fixed predictor reveals corpus-dependent changes in recognition performance and class predictions. 
The selected class transitions concentrate at particular concept levels, while on MELD predictions change despite little change in Macro-F1. 
These results show that aggregate scores can conceal changes in individual predictions. The concept bottleneck enables examination of  dependence through controlled changes to the predictor's inputs.

\clearpage
\section{Compliance with Ethical Standards}
This study uses existing data from the CREMA-D, IEMOCAP, and MELD datasets, with no new participant recruitment or data collection.
Ethical approval was not required for this secondary analysis.

\section{Acknowledgments}
The authors declare no conflicts of interest.
The authors used Claude Code and OpenAI Codex to assist with language editing and clarity improvements throughout the manuscript, as well as the development of experimental code.
The authors reviewed and verified the resulting revisions and code and take full responsibility for the final content.

\ninept
\bibliographystyle{IEEEbib}
\bibliography{refs}

\end{document}